\documentclass[twocolumn, 10pt, aps, prb, floatfix, showpacs, citeautoscript, superscriptaddress,notitlepage]{revtex4-1}

\usepackage{amsmath}

\usepackage[pdftex]{xcolor}
\usepackage{epsfig}
\usepackage{graphicx} 
\usepackage{bm} 
\usepackage{amssymb}
\usepackage{textcomp}
\usepackage{gensymb}
\usepackage{bbm}
\usepackage{comment}
\usepackage[colorlinks, citecolor={blue!50!black}, urlcolor={blue!50!black}, linkcolor={red!50!black}]{hyperref}
\newtheorem{lemma}{Lemma}
\DeclareMathOperator{\Tr}{Tr}

\newcommand{\co}[2]{#2}

\newcommand{\ket}[1]{\left| #1 \right\rangle}

\newcommand{\pars}[1]{\left( #1 \right)}
\newcommand{\brac}[1]{\left\{ #1 \right\}}

\newcommand{\br}{\mathbf{r}}
\newcommand{\bR}{\mathbf{R}}

\newcommand{\bk}{\mathbf{k}}

\newcommand{\bp}{\mathbf{p}}
\newcommand{\cH}{\mathcal{H}}
\newcommand{\cK}{\mathcal{K}}
\newcommand{\cP}{\mathcal{P}}
\newcommand{\cT}{\mathcal{T}}
\newcommand{\cC}{\mathcal{C}}
\newcommand{\id}{\mathbbm{1}}

\begin{document}
\bibliographystyle{apsrev4-1}
\title{Qsymm: Algorithmic symmetry finding and symmetric Hamiltonian generation}
\author{D{\'a}niel Varjas}
\email[Electronic address: ]{dvarjas@gmail.com}
\affiliation{QuTech, Delft University of Technology,
  P.O. Box 4056, 2600 GA Delft, The Netherlands}
\affiliation{Kavli Institute of Nanoscience, Delft University of Technology,
  P.O. Box 4056, 2600 GA Delft, The Netherlands}
\author{T{\'o}mas {\"O}. Rosdahl}
\email[Electronic address: ]{torosdahl@gmail.com}
\affiliation{Kavli Institute of Nanoscience, Delft University of Technology, P.O. Box 4056, 2600 GA Delft, The Netherlands}
\author{Anton R. Akhmerov}
\affiliation{Kavli Institute of Nanoscience, Delft University of Technology,
  P.O. Box 4056, 2600 GA Delft, The Netherlands}

\begin{abstract}
Symmetry is a guiding principle in physics that allows to generalize conclusions between many physical systems.
In the ongoing search for new topological phases of matter, symmetry plays a crucial role because it protects topological phases.
We address two converse questions relevant to the symmetry classification of systems:
Is it possible to generate all possible single-body Hamiltonians compatible with a given symmetry group?
Is it possible to find all the symmetries of a given family of Hamiltonians?
We present numerically stable, deterministic polynomial time algorithms to solve both of these problems.
Our treatment extends to all continuous or discrete symmetries of non-interacting lattice or continuum Hamiltonians.
We implement the algorithms in the Qsymm Python package, and demonstrate their usefulness with examples from active research areas in condensed matter physics, including Majorana wires and Kekule graphene.
\end{abstract}
\maketitle

\section{Introduction}
\co{Introduce important role of symmetries in condensed matter physics.}
A transformation that leaves a physical system invariant is called a symmetry, and such transformations have an ever-important role in modern physics.
For example, symmetry breaking characterizes the classical theory of phase transitions, and the invariance of the speed of light between reference frames is a cornerstone of special relativity theory.
In molecules and crystals, the symmetries of the constituent atomic orbitals determine the character of chemical bonds.
The band theory of solids uses the translational invariance of a crystal structure to classify states into energy bands according to their crystal momentum, where the band structure is in turn constrained by the point group symmetries of the crystal.
To describe such bands, model Hamiltonians based on tight-binding approximations\cite{Slater1954, Vogl1983, Jancu1998, Liu2013, Cappelluti2013} or $\bk \cdot \mathbf{p}$ perturbation theory\cite{Luttinger1955, Kormanyos2015} are typically constructed by fitting a generic Hamiltonian allowed by symmetry to match experimental data or first principles-calculations.

\co{Connection to symmetry-protected topological phases.}
Recent studies focused on the role of symmetry in protecting topological phases\cite{Hazan2010, Sato2015}.
Initially, analysis of time reversal and particle-hole symmetries led to the full classification of free fermionic phases in the ten Altland-Zirnbauer classes \cite{Altland1997, Schnyder2008, Kitaev2009}.
Later interest shifted to include symmetries involving transformations of space \cite{Fu2011, Ando2015, Po2017, Watanabe2017, Bradlyn2017, Bradlyn2018, Murakamie2017}.
Some of these phases are stable against disorder that preserves the symmetry only on average\cite{Fulga2014}, leading to a richer structure of symmetry-protected topological phases.
Analysis of newly proposed symmetry-protected topological phases is often done using minimal models, such as tight-binding Hamiltonians with short-range hoppings, or continuum Hamiltonians valid near high symmetry points in the Brillouin zone.
Although these models are usually easy to solve, they are prone to having higher symmetry than intended.
With the plethora of available symmetry groups, it is a nontrivial task to construct models that possess the stated, but only the stated symmetries, or to decide the complete symmetry group of a given Hamiltonian.

\begin{figure}[!tbh]
\includegraphics[width=0.97\columnwidth]{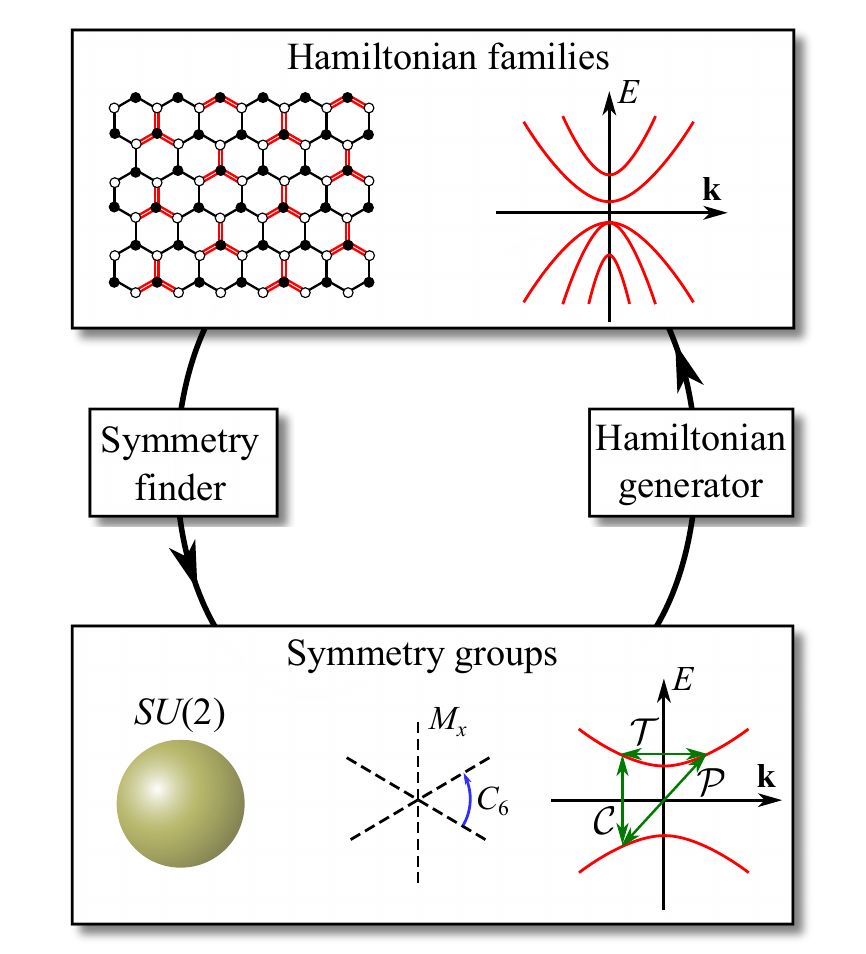}
\caption{Pictorial summary of the methods studied in the paper.
The symmetry finder and Hamiltonian generator algorithms form a two-way connection between Hamiltonian families and symmetry groups.}
\label{fig:summary}
\end{figure}
\co{Scope of methods}
In this paper we present an algorithm to generate all Hamiltonians that respect given symmetries, using an approach similar to Ref.~\onlinecite{kdotp-symmetry}.
In addition, we develop a dual algorithm to find all symmetries of a family of Hamiltonians (Fig.~\ref{fig:summary}).
Our framework is applicable to all non-interacting, finite or translation invariant lattice or $\bk \cdot \mathbf{p}$ Hamiltonians.
We treat all possible symmetries, including continuous unitary symmetry groups, continuous spatial rotations, space groups, discrete onsite symmetries (such as time reversal, particle-hole and chiral symmetries), and arbitrary combinations of these.
Besides static fermionic systems, our methods are also applicable to band structures in photonic crystals \cite{Notomi2000, Lu2014}, magnon spectra, classical mechanical \cite{Kane2013, Susstrunk2015} and electronic systems \cite{Ningyuan2015}, and driven Floquet systems\cite{Kitagawa2010}.

\co{Structure of paper.}
The paper is structured as follows: first we review the general symmetry structure of single-particle Hamiltonians.
Then we present our algorithm to find Hamiltonians with such symmetries.
After that we review the symmetry finding algorithm, which, by factoring out onsite unitary symmetries, makes finding all other symmetries more efficient and guaranteed.
We implement our algorithms in the Qsymm Python package \cite{supp_mat, supp_mat_2}.
Finally we provide a set of examples where our method was used on problems in active areas of research.
We show that Majorana wire devices may be protected against band-tilting by a magnetic symmetry, and double Dirac cones in Kekule distorted graphene are protected by point group and sublattice symmetry.
We also construct model Hamiltonians for transition metal dichalcogenides and distorted spin-orbit coupled SnTe.

\section{Hamiltonian families and symmetries}

\subsection{Continuum and tight-binding Hamiltonians}
\co{General framework of noninteracting Hamiltonians.}
We focus on non-interacting Hamiltonians.
The quadratic Hamilton operator of a finite (zero-dimensional) system can be written as
\begin{equation}
\hat{H} = \sum_{ij} H_{ij} \hat{a}^{\dag}_i \hat{a}_j,
\end{equation}
where $H$ is a Hermitian matrix and $\hat{a}_i$ are creation or annihilation operators.
We do not make any assumptions about bosonic or fermionic nature of these operators and also allow terms with two creation or two annihilation operators, facilitating the study of superconducting Bogoliubov-de Gennes Hamiltonians.
In the framework we use, all the details of the system are encoded in constraints on the matrix part $H$, which is the focus of our study.

\co{Translation invariance and k-space.}
Besides finite systems, we also address systems with $d$-dimensional translation invariance.
The associated conserved quantity is the (lattice) momentum $\bk$, which takes values in $\mathbbm{R}^d$ for continuous translations, and in the $d$-dimensional Brillouin zone for discrete translations.
Effective continuum Hamiltonians ($\bk \cdot \mathbf{p}$ models) also arise as the long-wavelength limit of lattice Hamiltonians.
The conservation of $\bk$ allows to decompose the single-particle Hilbert space into independent sectors corresponding to each $\bk$, such that $\hat{H}$ does not mix sectors, i.e.
\begin{equation}
\hat{H} = \sum_{ij} \sum_{\bk} H_{ij}(\bk) \hat{a}^{\dag}_{i\bk} \hat{a}_{j\bk}.
\end{equation}
In the rest of this work we focus on analyzing the matrix-valued Hamiltonian $H(\bk)$ with matrix elements $H_{ij}(\bk)$.

\co{tight-binding Hamiltonians.}
A tight-binding Hamiltonian acts on a Hilbert space consisting of basis orbitals in a single translational unit cell, and has the general form
\begin{equation} H(\bk) = \sum_{\bm{\delta}} \pars{e^{i \bk\cdot \bm{\delta}} h_{\bm{\delta}} + \textnormal{h.c.}}. \end{equation}
The hopping vectors $\bm{\delta}$ connect sites on the lattice, with the matrices of hopping amplitudes $h_{\bm{\delta}}$.
The $\bk \cdot \mathbf{p}$ Hamiltonian provides an accurate continuum approximation near a point in the Brillouin zone, typically a high symmetry point.
It is a polynomial in momenta $k_i$, and has the form
\begin{equation} H(\bk) =  \sum\limits_{\mathbf{n}} \bk^{\mathbf{n}} h_{\mathbf{n}}, \end{equation}
where $\bk^{\mathbf{n}} = \prod\limits_i  k_i^{n_i}$ is a monomial in the multi-index notation with $\mathbf{n} = (n_1, n_2, \hdots)$, and $h^{\dagger}_{\mathbf{n}} = h_{\mathbf{n}}$.
Typical methods to construct $\bk \cdot \mathbf{p}$ Hamiltonians start with a series expansion of a more complete lattice model from e.g.~\emph{ab initio} calculations around the high symmetry point, or by writing down all symmetry-allowed terms and fitting to experimental data or first principles calculations \cite{Vogl1983, Jancu1998, Fu2009, Xiao2012, Fang2015, Kormanyos2015}.

\subsection{Hamiltonian families}
\co{Hamiltonian families and why this is the right thing to consider.}
A set of symmetries only defines a \emph{Hamiltonian family}, as opposed to one single Hamiltonian.
A Hamiltonian family is the linear space of Hamiltonians
\begin{subequations}
\begin{equation}
\label{eq:family_a}
H(\bk) = \sum\limits_n c_n H_n(\bk),
\end{equation}
with arbitrary real coefficients $c_n$, and basis vectors
\begin{equation}
H_n (\bk) = \sum\limits_{\alpha} f_{\alpha} (\bk) h_{\alpha n},\;H_n ^\dagger (\bk) = H_n (\bk).
\label{eq:family_basis}
\end{equation}
\label{eq:family}
\end{subequations}
Here $h_{\alpha n}$ are constant matrices of identical size, and $f_{\alpha} (\bk)$ are linearly independent scalar functions.
In the rest of this work, whenever referring to a Hamiltonian, we mean a Hamiltonian family.
A Hamiltonian family is also the only useful starting point to analyzing the symmetry content of Hamiltonians.
For a zero-dimensional Hamiltonian, the symmetry group is always given by independent unitary transformations in each degenerate eigensubspace.
This group, however, provides no insight beyond the degeneracies of the levels.
In a family of continuum $\bk \cdot \mathbf{p}$ Hamiltonians, $f_{\alpha} (\bk)$ is a monomial, and in a tight-binding Hamiltonian, a phase factor $e^{i\bk\cdot \bm{\delta}}$ with $\bm{\delta}$ a hopping vector.

\co{Properties of the basis we use to expand k-space Hamiltonians.}
There is a natural inner product in the space of Hamiltonians \eqref{eq:family}.
We define the product of $H_1(\bk) = f_1(\bk) h_1$ and $H_2(\bk) = f_2(\bk) h_2$ as $\left< H_1(\bk), H_2(\bk) \right> = \left< f_1, f_2 \right> \left< h_1, h_2 \right>_F$.
On the matrix part, the Frobenius inner product is given by $\left< A, B \right>_F = \Tr\pars{A^{\dag} B}$.
For the inner product of the $\bk$-dependent prefactors, we use the Bombieri inner product\cite{Bombieri1990} for polynomials, such that different monomials are orthogonal, and $\left< e^{i \bk\cdot\mathbf{a}}, e^{i \bk\cdot\mathbf{b}} \right> = \delta_{\mathbf{a}, \mathbf{b}}$ for phase factors.
Both of these inner products on the function spaces are invariant under the isometries of $\bk$-space, and therefore all symmetry actions we consider in this work are (anti)unitary with respect to this inner product.
This structure of the space of Hamiltonians also justifies our use of single exponentials and single monomials as the expansion basis.

\subsection{Symmetry constraints on Hamiltonian families}
\label{sec:sym_constraints}
\co{General properties of symmetry transformations.}
We adopt the active view of symmetry action $g$ on the Hamiltonian: $g(H)$ represents a transformed Hamiltonian, such that the matrix elements between rotated wave functions $g\pars{\ket{\psi(\bk)}}$ are identical.
In other words, a Hamiltonian has a symmetry if $g$ leaves the Hamiltonian invariant,
\begin{equation}
g(H) = H \label{eq:trans}.
\end{equation}
A general unitary symmetry $g$ acts on a Hamiltonian $H(\bk)$ as
\begin{subequations}
\label{eqn:trfU}
\begin{equation}
g\pars{H}(\bk) = \pm U_g H( R_g^{-1} \bk) U_g^{-1},
\end{equation}
and a general antiunitary symmetry as
\begin{equation}
g\pars{H}(\bk) = \pm U_g H^*( -R_g^{-1} \bk) U_g^{-1}.
\end{equation}
\end{subequations}
Here the orthogonal matrix $R_g$ is the real space action, and the unitary matrix $U_g$ is the Hilbert space action of $g$.
We include the overall $\pm$ sign to treat \emph{antisymmetries}---symmetries that reverse the sign of energy---on an equal footing.
We restrict our considerations to a constant $U_g$, however, in the real space basis (see Section \ref{section:gen_lattice}), any space group operator may only contain an overall $\bk$-dependent phase factor, which cancels in the previous equations.

\co{Formal properties, action of conjugation on k.}
Substituting Eq.~(\ref{eqn:trfU}) into (\ref{eq:trans}), we rewrite it in a form linear in the symmetry action:
\begin{equation}
\mathcal{S} H(\bk) = \pm H( R \bk) \mathcal{S} \label{eq:constraint}.
\end{equation}
Here the symmetry action $\mathcal{S}$ is $\mathcal{S} = U$ if unitary and $\mathcal{S} = U \cK$ if antiunitary, with $\cK$ the real space complex conjugation operator: $\cK H(\bk) \cK = H^*(-\bk)$.

\co{Symmetry constraints on Hamiltonian families.}
We apply the symmetry constraint \eqref{eq:constraint} to the Hamiltonian family \eqref{eq:family_a}.
The spatial action of a symmetry is a rotation in the space of $f_{\alpha}$, such that $f_\alpha (\pm R \bk) = \sum_{\beta} \gamma_{\alpha}^{\beta} f_{\beta} (\bk)$ with $\gamma_{\alpha}^{\beta}$ known for given $R$ and $f_{\alpha}$.
Substituting this yields
\begin{equation}
\sum\limits_\alpha f_\alpha (\bk) \left[ \sum\limits_n (  \mathcal{S} h_{\alpha n} \mp \sum\limits_\beta \gamma_\beta^\alpha h_{\beta n} \mathcal{S} ) c_n \right] = 0.
\end{equation}
Since $f_{\alpha}$ are linearly independent functions, the matrix coefficients in the parentheses must vanish for every $\alpha$, resulting in the system of equations
\begin{equation}
\sum\limits_n (  \mathcal{S} h_{\alpha n} \mp \sum\limits_\beta \gamma_\beta^\alpha h_{\beta n} \mathcal{S} ) c_n = 0,~\forall \alpha. \label{eq:linear_constraints}
\end{equation}

\co{Continuous families of symmetries.}
When symmetries form continuous Lie groups, it is advantageous to use the symmetry generators instead of the group elements.
Consider a one parameter family of transformations $g_{\phi}$ which, for a fixed $\phi$ act as a unitary symmetry in the above, and let $g_0$ be the identity.
We define the action of the generator $g'$ through
\begin{equation}
g'(H) = \left.\frac{d}{d\phi} g_{\phi}(H)\right|_{\phi = 0}.
\end{equation}
Substituting (\ref{eqn:trfU}) and using that $U_{g_{\phi}} = e^{- i \phi L}$ with $L = L^{\dag}$ and $R_{g_{\phi}} = e^{-i \phi M}$ with $M = -M^T = M^{\dag}$, we find
\begin{equation}
g'(H)(\bk) = i [H(\bk), L] + i \sum_{ij}\frac{\partial H}{\partial k_i} M_{ij} k_j.
\end{equation}
$g_{\phi}(H) = H$ for every $\phi$ is equivalent to $g'(H) = 0$.
Tight-binding Hamiltonians cannot be invariant under continuous rotations of space, such that $M = 0$ and the symmetry constraint simplifies to
\begin{equation}
\sum_n [h_{\alpha n}, L] c_n = 0, \label{eq:tb_continuous}
\end{equation}
where $L$ is a local conserved quantity.
Finally, if the Hamiltonian is a polynomial in $\bk$, continuous rotation invariance is also possible.
With $\sum_{ij} \frac{\partial f_\alpha}{\partial k_i} M_{ij} k_j = \sum_{\beta} \gamma_{\alpha}^{\beta} f_{\beta} (\bk)$ the symmetry constraint reads
\begin{equation}
\sum\limits_n ([h_{\alpha n}, L] + \sum_{\beta} \gamma_{\alpha}^{\beta} h_{\beta n} ) c_n = 0. \label{eq:kp_continuous}
\end{equation}

\section{Generating Hamiltonians from symmetry constraints}
\subsection{Constraining Hamiltonian families}

\co{General formulation of the problem.}
Given a symmetry $\mathcal{S}$ and a Hamiltonian family \eqref{eq:family}, we wish to find the subfamily of Hamiltonians that is invariant under the symmetry transformation \eqref{eq:constraint}.
The symmetry constraint on the Hamiltonian family is a system of homogeneous linear equations for the coefficients $c_n$ [see Eqs.~\eqref{eq:linear_constraints}, \eqref{eq:tb_continuous}, and \eqref{eq:kp_continuous}].
We find the space of solutions numerically using singular value decomposition or sparse eigendecomposition, which gives the subfamily of the original Hamiltonian family \eqref{eq:family} that satisfies the symmetry.
Imposing additional symmetry constraints on the family yields further linear equations that are identical to Eq.~\eqref{eq:linear_constraints} in form.
We provide an implementation of this algorithm in the Qsymm Python package.

\co{Hexagonal warping example.}
The constraining algorithm allows to generate all possible tight-binding or $\bk \cdot \mathbf{p}$ Hamiltonians that satisfy symmetry constraints, by applying the algorithm to the most general representative Hamiltonian family for the system at hand.
As an illustration, we reproduce the family of two-band $\bk \cdot \mathbf{p}$ Hamiltonians of Ref.~\onlinecite{Fu2009} for the surface dispersion of the topological insulator Bi$_2$Te$_3$.
Our starting point is the family of all $2\times 2$ $\bk \cdot \mathbf{p}$ Hamiltonians up to third order in the momentum ${\bf k} = (k_x, k_y)$.
Expanding the matrix part in terms of the identity and Pauli matrices $\sigma_{0, x, y, z}$, the general family consists of $40$ basis vectors and is given by
\begin{equation} H({\bf k}) = \sum\limits_{\bf n} c_{\bf n} \sigma_j k_x^{\alpha_x} k_y^{\alpha_y},~0\leq|\alpha|\leq 3, \label{eq:general_fam} \end{equation}
with ${\bf n} = (j, \alpha_x, \alpha_y) = (j, \alpha)$.
To obtain the surface dispersion Hamiltonian, we constrain \eqref{eq:general_fam} with time-reversal symmetry ($\mathcal{T} = i\sigma_y K$), and the point group symmetries of the crystal, namely three-fold rotation ($C_3 = e^{-i\pi \sigma_z/3}$), and mirror symmetry in $x$ ($M=i\sigma_x$) \cite{Fu2009}.
Substituting the three symmetries and the family \eqref{eq:general_fam} into \eqref{eq:linear_constraints} yields a homogeneous system of $120$ linear equations for the $40$ coefficients $c_{\bf n}$.
The null space of the linear system is the subfamily of Hamiltonians that satisfy the symmetry constraints:
\begin{equation}
\begin{split}
H({\bf k}) = & c_0 k^2 \sigma_0 + c_1(k_x \sigma_y - k_y \sigma_x) + c_2(k_x^3 - 3k_x k_y^2) \\
 & + c_3(k_x k^2 \sigma_y - k_y k^2 \sigma_x),
\end{split}
\end{equation}
with $c_n \in \mathbb{R}$, which matches the Hamiltonian of Ref.~\onlinecite{Fu2009}.
Here, $k^2 = k_x^2 + k_y^2$, and we have relabelled the coefficients $c_n$ for clarity.

\subsection{Generating lattice Hamiltonians by symmetrization}
\label{section:gen_lattice}
\co{Reason to use other method.}
Lattice models often contain multiple sites per unit cell, but only a small number of bonds.
In this case the previous approach of generating all possible terms and constraining them is inefficient due to the large dimension of the null-space.
On the other hand, all the hopping terms on symmetry equivalent bonds are completely determined by the hopping on one of these bonds.
This allows us introduce a symmetrization strategy to generate all symmetry-constrained lattice Hamiltonians with hoppings of limited range.

\co{Basis for tight-binding Hamiltonians on arbitrary lattice.}
To treat arbitrary space group symmetries of general crystal structures, we consider tight-binding Hamiltonians in the real space basis that preserves information on the coordinates of the basis orbitals\cite{Resta2000, Varjas2015}.
Up to a normalization factor, the Bloch basis functions are given by $\ket{\chi_{\bk}^{al}} = \sum_{\bR} e^{i \bk (\bR+\mathbf{r}_a)} \ket{\phi_{\bR}^{al}}$, where $a$ indexes the sites in the unit cell, $l \in [1, \ldots, n_{\textnormal{orbs.}}^a]$ indexes the orbitals on the site, $\br_a$ is the real space position of the site and $\bR$ runs over all lattice vectors.
In this basis, the hopping terms in the Hamiltonian acquire a phase factor corresponding to the true real space separation of the sites they connect, as opposed to the separation of the unit cells to which the sites belong.
A hopping between site $a$ at $\br_a$ and site $b$ at $\br_b = \br_a + \bm{\delta}_{ab}$ enters as a term $e^{i\bk\bm{\delta_{ab}}} h^{ab}_{\bm{\delta}_{ab}} + \textnormal{h.c.}$ where we suppressed the orbital indices of the matrix $h^{ab}_{\bm{\delta}_{ab}}$.
Onsite terms have $\bm{\delta}_{aa} = 0$.
The main advantage of using this gauge is that the form of the Hamiltonian is independent of the choice of the real space origin and the shape of the unit cell.
As a consequence, nonsymmorphic symmetry operations only acquire $\bk$-dependence in the form of an overall phase factor\cite{Varjas2015}.
In the simplest case of a single site per unit cell, $\bm{\delta}_{ab}$ are lattice vectors.

\co{Symmetrization strategy for generating Hamiltonian.}
We start from a small set of terms for every symmetry unique bond $\bm{\delta}_{ab}$ of the form
\begin{equation}
H_n(\bk) = e^{i\bk\bm{\delta}_{ab}} h^{ab}_{n\bm{\delta}_{ab}} + \textnormal{h.c.},
\end{equation}
with $h^{ab}_{n\bm{\delta}_{ab}}$ spanning all $n_{\textnormal{orbs.}}^a \times n_{\textnormal{orbs.}}^b$ matrices that are invariant under the continuous onsite symmetry group.
We symmetrize these with respect to the discrete point group $G$, i.e.~
\begin{equation}
H_s = \frac{1}{|G|}\sum_{g\in G} g(H),
\end{equation}
where $g(H)$ is the symmetry transformed image of $H$ under the transformation $g$ [see Eq.~\eqref{eq:trans}] and $|G|$ is the number of elements in $G$.
Because the sum over $g\in G$ can be replaced by a sum over $(hg) \in G$, $h(H_s) = H_s$ for all $h\in G$.
In addition, $H_s$ is exactly the projection of $H$ onto the space of symmetric Hamiltonians.
The symmetrized terms span all symmetry allowed Hamiltonians with the prescribed hopping vectors.
This space of Hamiltonians is generally overcomplete, we find a minimal set of terms spanning the space using standard linear algebra techniques.

\co{Graphene example}
As an example, consider graphene with one spinless orbital per site.
A three-fold rotation around a site maps both sublattices onto themselves, so the unitary part of the symmetry action is $\id_{2\times 2}$. Let $\bm{\delta} = (a_0, 0)$ be a vector connecting nearest neighbors, and the corresponding hopping term
\begin{equation}
H = t \begin{bmatrix}
0 & e^{i \bk\cdot\bm{\delta}} \\
e^{-i \bk\cdot\bm{\delta}} & 0
\end{bmatrix} =
t \begin{bmatrix}
0 & e^{i a_0 k_x} \\
e^{-i a_0 k_x} & 0
\end{bmatrix},
\end{equation}
which is Hermitian and only connects the two sublattices.
After symmetrization with respect to the full hexagonal group we obtain the well known minimal tight-binding model for graphene:
\begin{equation}
H_s = \frac{t}{3} \begin{bmatrix}
0 & h \\
h^{\dag} & 0
\end{bmatrix}
\end{equation}
with $h = e^{i a_0 k_x} + e^{i a_0 \left(-\frac{1}{2} k_x + \frac{\sqrt{3}}{2} k_y \right)} + e^{i a_0 \left(-\frac{1}{2} k_x - \frac{\sqrt{3}}{2} k_y \right)}$.

\section{Symmetry finding}
\begin{figure}[!tbh]
\includegraphics[width=0.97\columnwidth]{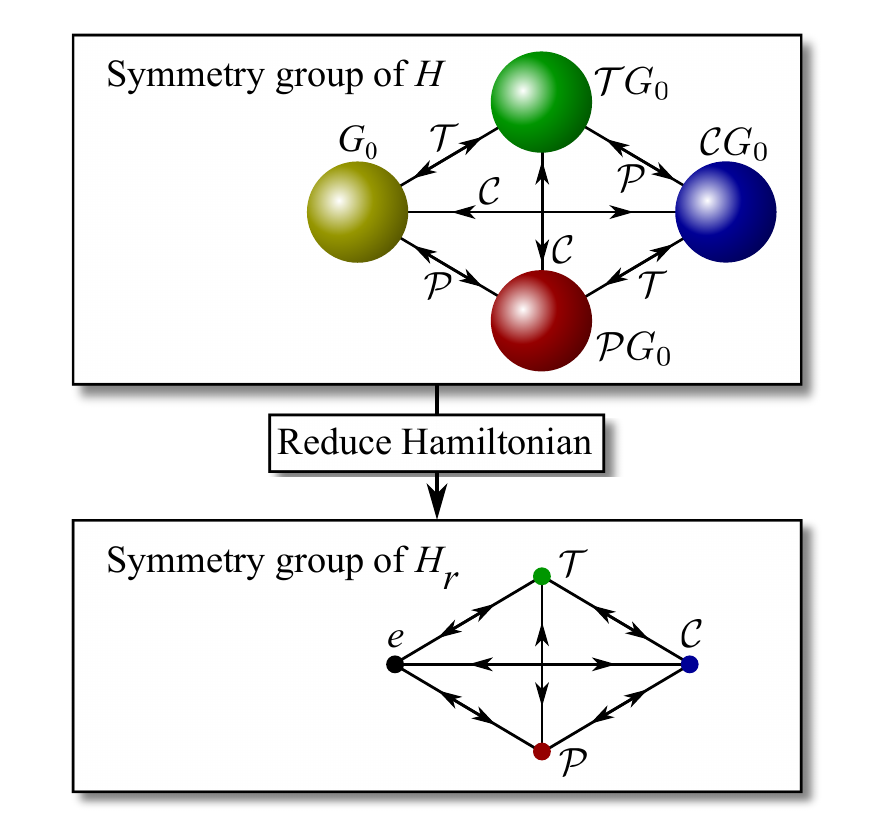}
\caption{Schematic representation of the onsite symmetry group of a Hamiltonian family (top). The unitary symmetries form a continuous connected Lie group $G_0$. The discrete symmetries can all be combined with any unitary symmetry, forming disconnected components of the symmetry group. For example, $\cT G_0 = \brac{\cT U : U \in G_0}$ contains all antiunitary symmetries which are combinations of the canonical time reversal $\cT$ and some unitary $U$. Reducing the Hamiltonian family factors out all the unitary symmetries leaving only the identity element $e$, resulting in a simpler discrete group structure (bottom).}
\label{fig:groups}
\end{figure}

\co{Summarize section.}
Unlike finding a family of symmetric Hamiltonians, that amounts to solving a linear system, finding the symmetries of a Hamiltonian family is more involved.
We first focus on finite (zero-dimensional) systems and show that the unitary symmetry group generally admits a continuous Lie group structure.
Next we present an algorithm to find the unitary symmetries, and rewrite the Hamiltonian family in the symmetry-adapted basis.
In this basis the Hamiltonian takes a block diagonal form, where the blocks are guaranteed to have no unitary symmetries, hence we call these blocks reduced Hamiltonians.
Factoring out the unitary symmetries this way simplifies finding the discrete (anti)unitary (anti)symmetries, see Fig.~\ref{fig:groups}.
After generalizing these methods to onsite symmetries of translation invariant systems in arbitrary dimensions, we finally include real space rotation symmetries.
\subsection{Structure of the onsite unitary symmetry group}
\co{Onsite unitary symmetry implies conserved quantity and continuous symmetry group.}
Assume a unitary symmetry operator $U$ commutes with a family of finite Hamiltonians:
\begin{equation}
[U, H_n] = 0.
\label{eqn:commU}
\end{equation}
Any unitary is expressible as the exponential of a Hermitian operator $L$
\begin{equation}
U = e^{- i L},
\end{equation}
which is unique if we restrict the spectrum $\sigma(L) \subset [0 , 2\pi)$ (this condition is equivalent to choosing a branch cut for the logarithm in $L = i \log U$).
Defining $U$ and $L$ this way ensures that they have exactly the same eigensubspaces.
Since $U$ and $H_n$ commute, they also share eigensubspaces, and Eq.~\eqref{eqn:commU} is equivalent to
\begin{equation}
[L, H_n] = 0.
\end{equation}
Therefore the Hamiltonian family has a continuous family of unitary symmetries $U(\phi) = e^{- i \phi L}$ that all commute with $H_n$.
Because a single unitary symmetry defines a continuous family that is connected to the identity, the full group of unitary symmetries must form a single connected component $G_0$.
This is uniquely specified by the space of conserved quantities $L \in \mathfrak{g}$, the Lie algebra of the Lie group $G_0$.

\co{Example with spin conservation.}
Consider for example a system consisting of a number of spinful orbitals, which is invariant under the set of unitary spin-flip operators $U_i = i \sigma_i \otimes \id$, where $\sigma_i$ are the Pauli matrices and the identity acts on the space of orbitals.
Taking the logarithms of these operators, we find that they are associated with the Hermitian conserved spins $L_i = \sigma_i \otimes \id$.
The Lie group generated by these conserved quantities is $SU(2)$ acting in spin space.
Therefore the generic Hamiltonian of such a system assumes the form $H = \id_{2\times 2} \otimes H_r$, where the reduced Hamiltonian $H_r$ acts only on the space of orbitals and the identity acts on spin space.
In the basis where spin up and spin down states are grouped together, the original Hamiltonian takes the block diagonal form with two identical blocks $H = H_r \oplus H_r$, reducing the problem to spinless fermions.

\co{Example with higher spin conservation.}
Consider the same system, but with higher spins on every site instead.
Let the conserved spins be $L_i = J_i \otimes \id$ where $J_i$ for $i \in [x, y, z]$ form a $2s + 1$ dimensional spin representation.
The generic Hamiltonian again has the form $H = \id_{(2s + 1)\times (2s + 1)} \otimes H_r$, because the $J_i$ form an irreducible representation of the rotation group.
This Hamiltonian, however, is invariant under any unitary transformation of the form $U\otimes\id$ with $U \in U(2s +1)$.
We therefore find that the symmetry group is in fact larger than the one we started with, forming a full unitary group.

\co{Why this only applies to non-interacting systems.}
The above result may sound surprising on physical grounds, considering that many well-studied models (\emph{e.g.} the transverse field Ising model) only have discrete onsite symmetry groups.
We emphasize that this result (and much of what follows) is specific to single-particle systems, and does not directly apply to onsite symmetries on the many-particle Fock space.
In the single-particle case the full Hilbert space is the direct sum of the local Hilbert spaces and therefore an onsite symmetry is the direct sum of local unitaries.
The many-particle Hilbert space is a direct product, and onsite unitary symmetries take a direct product form.
The associated $L$ is generally not a sum of local terms, and does not correspond to a local conserved quantity.
The above argument also fails when considering spatial symmetries, because in general the logarithm of a locality-preserving operator (an operator that maps a state with localized support to one with localized support) mixes degrees of freedom that are far apart.

\co{In the following there will be a lot of math, you can jump ahead.}
In the rest of this subsection we prove, using the theory of Lie groups, that the unitary symmetry group $G_0$ is a direct product of unitary groups $U(N)$ in any finite system.
We then show the existence of the symmetry-adapted basis, where both the conserved quantities and the Hamiltonian take a simple form, and derive properties of reduced Hamiltonians.
The reader not interested in mathematical proofs may skip to the next subsection where the algorithm for finding unitary symmetries is discussed.

\co{Onsite Unitary symmetry group is a subgroup of U(n), with nice properties.}
The unitary symmetry group $G_0$ is a subgroup of the full unitary group on the Hilbert space $\mathcal{H}$, $G_0 \leq U(\dim \cH)$.
$G_0$ is a connected and compact matrix Lie group, which means that all of its finite-dimensional representations are completely reducible\cite{hall2015, Fulton1991}.
The Lie group $G_0$ is generated by all the generators in its Lie algebra $L\in\mathfrak{g}$ for which $[L, H] = 0$, the Lie algebra $\mathfrak{g}$ is also completely reducible.

\co{Irreducible subspaces.}
Reducing the representation amounts to splitting the Hilbert space $\cH$ into a direct sum of irreducible subspaces $\mathcal{V}^{(i)}_j$:
\begin{equation}
\cH = \pars{\mathcal{V}^{(1)}_1 \oplus \ldots \oplus \mathcal{V}^{(1)}_{n_1}} \oplus  \pars{\mathcal{V}^{(2)}_1 \oplus \ldots \oplus \mathcal{V}^{(2)}_{n_2}} \oplus \ldots .
 \label{eqn:blockV}
\end{equation}
Each of these subspaces is invariant under the symmetry action and contains no invariant subspace.
$\mathcal{V}^{(i)}_j$ transforms according to irreducible representation (irrep) $i$, and irrep $i$ has multiplicity $n_i$.
We denote the union of all irreducible subspaces belonging to irrep $i$ as $\mathcal{V}^{(i)} = \mathcal{V}^{(i)}_1 \oplus \ldots \oplus \mathcal{V}^{(i)}_{n_i}$.

\co{symmetry-adapted basis and block diagonal form of symmetries.}
In a \emph{symmetry-adapted basis}, every symmetry generator takes the same block diagonal form of irreducible representations:
\begin{eqnarray}
L &=& \underbrace{L^{(1)} \oplus L^{(1)} \oplus \ldots}_{n_1 \textnormal{times}} \oplus  \underbrace{L^{(2)} \oplus L^{(2)} \oplus \ldots}_{n_2 \textnormal{times}} \oplus \ldots =\nonumber\\
&=& \pars{L^{(1)}  \otimes \id_{n_1\times n_1} } \oplus \pars{L^{(2)} \otimes \id_{n_2\times n_2} } \oplus \ldots,
 \label{eqn:blockL}
\end{eqnarray}
where $L^{(i)}$ is the representation of $L$ in the $i$-th irrep, each acting in a corresponding irreducible subspace $\mathcal{V}^{(i)}_j$ of dimension $d_i$.
Irreps $j$ and $k$ are equivalent if there exists a unitary transformation $W$ such that $L^{(j)} = W L^{(k)} W^{-1} \;\forall L\in \mathfrak{g}$.
This guarantees that there is a basis where all operators have exactly the same representation in every equivalent irreducible subspace.

\co{Show that H is also block diagonal and reduced Hamiltonians have no symmetry.}
In this basis, the Hamiltonian also takes a simple block form.
By Schur's Lemma, blocks of $H$ between irreducible subspaces that transform according to different irreps are zero, and blocks between irreducible subspaces with identical irreps are proportional to the identity:
\begin{equation}
H = \pars{\id_{d_1 \times d_1} \otimes H_1} \oplus \pars{\id_{d_2 \times d_2} \otimes H_2} \oplus \ldots,
\label{eqn:blockH}
\end{equation}
where the \emph{reduced Hamiltonians} $H_i$ are $n_i \times n_i$ Hermitian matrices.
The reduced Hamiltonians $H_i$ cannot have any nontrivial unitary symmetries.
To prove that, assume that $H_1$ has a conserved quantity $L$ such that $[H_1, L] = 0$.
It implies that $\pars{\id_{d_1 \times d_1} \otimes L} \oplus \,0\, \oplus \ldots$ commutes with the full Hamiltonian, which is incompatible with the unique decomposition to irreducible subspaces, except the trivial case of $L\propto\id$.

\co{Show that the full symmetry group is a product of U(n)'s.}
It is apparent from (\ref{eqn:blockH}) that the symmetry group of $H$ is a product of full unitary groups acting independently on each block
\begin{equation}
G_0 = U(d_1) \times U(d_2) \times\ldots,
\end{equation}
where the symmetry generators have the form (\ref{eqn:blockL}), with $L^{(i)} \in \mathfrak{u}(d_i)$ independently running over all $d_i \times d_i$ Hermitian matrices, and $\mathfrak{u}(d_i)$ the Lie algebra of $U(d_i)$.
Because the reduction to irreducible subspaces is unique, this is the full group of unitary symmetries.
The center of the group $Z(G_0)$ is formed by the abelian $U(1)$ subgroups generated by the set of projectors on each block, \emph{i.e.} generators where one of the $L^{(i)}$ is the identity and the others vanish.

\co{Characterize center through Killing form}
To have a basis-independent characterization of the center of the Lie algebra, we compute the structure constants ${f_{\alpha\beta}}^{\gamma}$ defined by
\begin{equation}
[L_{\alpha} , L_{\beta}] = i \sum_{\gamma} {f_{\alpha\beta}}^{\gamma} L_{\gamma}.
\end{equation}
Using these we define the Killing form
\begin{equation}
K_{\alpha\beta} =  \sum_{\gamma\delta} {f_{\alpha\gamma}}^{\delta}  {f_{\beta\delta}}^{\gamma}.
\end{equation}
It can be shown\cite{Fulton1991} that the null-space of the Killing form is exactly the center of the Lie algebra, \emph{i.e.} if a vector $l$ is a solution of $\sum_{\beta} K_{\alpha\beta} l^{\beta} = 0$ then $\sum_{\alpha} l^{\alpha} L_{\alpha}$ commutes with every operator in $\mathfrak{g}$.

\subsection{Finding the unitary symmetry group}
\label{sec:find_U}

\co{Algorithm to find all generators and center of the group.}
We are now ready to define the algorithm of finding the unitary symmetry group and constructing the reduced Hamiltonians for a given family of Hamiltonians.
First we find all symmetry generators $L_\alpha$ as the linearly independent solutions of
\begin{equation}
[L_{\alpha}, H] = 0 \;\textnormal{and}\; L_{\alpha} = L_{\alpha}^{\dag}.
\end{equation}
This is a system of linear equations for the unknown components of $L_{\alpha}$, which we solve using the same methods we used for constraining Hamiltonians.
After computing the Killing form $K$ we find all linearly independent solutions of
\begin{equation}
\sum_{\beta} K_{\alpha\beta} l^{\beta} = 0
\end{equation}
for $l$.
Operators of the form $\sum_{\alpha} l^{\alpha} L_{\alpha}$ are the basis of conserved quantities that commute with every other conserved quantity.
We simultaneously diagonalize all of these (see Appendix~\ref{sec:simult_diag}) to find the simultaneous eigensubspaces $\mathcal{V}^{(i)}$ (\ref{eqn:blockV}).

\co{Finding symmetry-adapted basis in a block.}
We then find the generators $L^{(i)}$ of the $SU(d_i)$ symmetry group of each block.
To do so, we project the Hamiltonian onto $\mathcal{V}^{(i)}$ using the projector $P_i$, which is an orthonormal set of column vectors, and solve
\begin{equation}
[L^{(i)}, P_i^{\dag} H P_i] = 0 \;\textnormal{and}\; L^{(i)} = {L^{(i)}}^{\dag} \;\textnormal{and}\; \Tr L^{(i)} = 0,
\end{equation}
to find the $d_i^2 - 1$ linearly independent solutions for $L^{(i)}$. 
The final step is finding a basis within $\mathcal{V}^{(i)}$ that gives the tensor product structure of (\ref{eqn:blockL}) and (\ref{eqn:blockH})
(see Appendix~\ref{sec:sym_bas}).
We use this basis and the resulting reduced Hamiltonians in the following.

\subsection{Discrete onsite symmetries and antisymmetries}
\label{sec:find_disc}

\co{Summarize strategy using reduced Hamiltonians.}
Next we discuss the \emph{discrete onsite symmetries}:
\begin{itemize}
\item time reversal (antiunitary symmetries),
\item particle-hole (antiunititary antisymmetries),
\item chiral (unitary antisymmetries).
\end{itemize}
These symmetries also form continuous families, because combining them with any onsite unitary symmetry also results in a discrete onsite symmetry of the same type.
Because there is no continuous way to interpolate between unitary and antiunitary symmetries or between symmetries and antisymmetries, each type forms a disconnected component of the onsite symmetry group (see Fig.~\ref{fig:groups}).
To find one representative of each type of the discrete onsite symmetries, we utilize the symmetry-adapted basis and reduced the Hamiltonian found in the previous subsection.
The reduced Hamiltonians have no residual symmetries, which makes the discrete onsite symmetries unique and allows us to efficiently find them.

\co{TR type symmetries, naive constraint.}
We start with time reversal symmetries of finite (zero-dimensional) systems.
$\cT$ is a time reversal if
\begin{equation}
\cT H = H \cT.
\end{equation}
Writing $\cT = U \cK$ with unitary $U$ and complex conjugation $\cK$, we obtain
\begin{subequations}
\begin{eqnarray}
\label{eq:Tfull}
U H^* &=& H U, \\
\label{eq:unitarity}
U U^{\dag} &=& \id.
\end{eqnarray}
\end{subequations}
This is a nonlinear system of equations, and it is in general hard to solve.
We show that using the reduced Hamiltonian simplifies it to a linear problem.

\co{Case of a single block in H.}
We first consider a Hamiltonian that has one set of identical irreducible subspaces, \emph{i.e.} $H = \id \otimes H_1$.
By (\ref{eqn:blockL}), all conserved quantities have the form $L =  L^{(1)} \otimes \id$, and span the full space of Hermitian matrices on the first Hilbert space of the tensor product.
If $L$ is a conserved quantity, so is $\cT L \cT^{-1}$, which implies $U \mathfrak{g} U^{\dag} = \mathfrak{g}$.
Therefore the unitary part of $\cT$ is a direct product of two unitaries, $U = V \otimes W$ (see Appendix \ref{app:tensor}), with $V$ an arbitrary unitary matrix.
Because $V = \id$ commutes with all unitary symmetries, we call $\cT = \id \otimes W \cK$ the canonical time reversal symmetry.
Due to the tensor product structure of $\cT$, \eqref{eq:Tfull} reduces to
\begin{equation}
W H_1^* = H_1 W.
\label{eqn:Tdiag}
\end{equation}
Importantly $H_1$ has no unitary symmetries.
Any nonzero solution $W$ of \eqref{eqn:Tdiag} has $\ker W = 0$: otherwise either $\ker W$ is an invariant subspace of $H_1^*$ or $\ker H_1^*$ is nonzero, both incompatible with $H_1$ having no unitary symmetries.
Considering two solutions $W$ and $\tilde{W}$ of \eqref{eqn:Tdiag} we find
\begin{equation}
W \tilde{W}^{\dag} H_1 = W H_1^* \tilde{W}^{\dag}  = H_1 W \tilde{W}^{\dag}.
\end{equation}
Because $H_1$ has no unitary symmetries $W \tilde{W}^{\dag} \propto \id$, which proves that any solution of \eqref{eqn:Tdiag} is unique and unitary up to a constant factor.
In other words, any normalized solution of \eqref{eqn:Tdiag} automatically satisfies \eqref{eq:unitarity}.

\co{Two simple examples.}
As an example consider the reduced family of Hamiltonians
\begin{equation}
H = c_1 \sigma_x + c_3 \sigma_z.
\end{equation}
This family has no residual symmetry, and solving (\ref{eqn:Tdiag}) we find that $W$ has to commute with both $\sigma_x$ and $\sigma_z$, so $W\propto \sigma_0$.
Therefore $H$ is invariant under $\cT = \sigma_0 \cK$, which is unique up to a phase factor.
As a second example consider the reduced family
\begin{equation}
\begin{split}
H =~& c_{10} \tau_x \sigma_0 + c_{30} \tau_z \sigma_0 + c_{21}\tau_y \sigma_x + \\
& c_{22}\tau_y \sigma_y + c_{23}\tau_y \sigma_z,
\end{split}
\end{equation}
with $\tau_i$ the Pauli matrices.
The condition (\ref{eqn:Tdiag}) implies that $W$ commutes with $\tau_x \sigma_0$, $\tau_z \sigma_0$, $\tau_y \sigma_y$ and anticommutes with $\tau_y \sigma_x$, $\tau_y \sigma_z$.
The only solution is $W\propto \tau_0 \sigma_y$, and $\cT = \tau_0 \sigma_y \cK$ up to a phase factor.

\co{General case with multiple blocks.}
In the general case of multiple irreps, we find (see Appendix~\ref{sec:disc_block} for details) that a time reversal can only mix subspaces $\mathcal{V}^{(i)}$ and $\mathcal{V}^{(j)}$ if they correspond to irreps of the same dimensionality and multiplicity.
The block structure of $U$ has to be symmetric, it can only exchange subspaces pairwise or leave subspaces invariant.
In order to find a time reversal, we iterate over all symmetric permutations of compatible subspaces, and check if a time reversal exists with the given block structure.
Specifically we consider two reduced blocks of the Hamiltonian that are interchanged
\begin{equation}
H_r = \begin{bmatrix}
H_i   & 0    \\
 0     & H_j
\end{bmatrix}.
\end{equation}

Because $H_i$ and $H_j$ have no unitary symmetry, the square of time reversal must have the form
\begin{equation}
\cT_r^2 = \begin{bmatrix}
e^{i\phi}\id   & 0    \\
 0     & e^{i\phi'}\id
\end{bmatrix}.
\end{equation}
Therefore, following Appendix~\ref{sec:disc_block} we search for a time reversal of the form
\begin{equation}
\cT_r = \begin{bmatrix}
 0   & W_{ij}    \\
 e^{i\phi} W_{ij}^T     & 0
\end{bmatrix} \cK.
\end{equation}
The relation $\cT_r H_r = H_r \cT_r$ then reduces to
\begin{equation}
W_{ij} H_j^{*} = H_i W_{ij}.
\label{eqn:Toffdiag}
\end{equation}
This is the key result of this section, and it is a generalization of \eqref{eqn:Tdiag}.
Following the same reasoning as before, we conclude that any nonzero solution $W_{ij}$ of \eqref{eqn:Toffdiag} is unique and unitary up to a constant factor.

\co{A more complicated example.}
As an example of this case consider the Hamiltonian family
\begin{equation}
H = \begin{bmatrix}
c_{10} \tau_x \!+\! c_{30} \tau_z \!+\! c_{23}\tau_y  & 0    \\
 0     &c_{10} \tau_x \!+\! c_{30} \tau_z \!-\! c_{23}\tau_y
\end{bmatrix}.
\end{equation}
Solving (\ref{eqn:Toffdiag}) for $W_{12}$ we find that $[W_{12}, \tau_i] = 0$ for $i=x,y,z$, so $W_{12} \propto \tau_0$ and $\cT = \tau_0 \sigma_x  \cK$ up to a phase factor.
We also confirm that (\ref{eqn:Toffdiag}) does not have any solutions for $W_{11}$ or $W_{22}$, so the block form of $\cT$ is unique.

\co{Extension to chiral and PH.}
Likewise, the canonical unitary and antiunitary antisymmetries act as a tensor product $U_{ij} = \id \otimes W_{ij}$ in each block, and either leave subspaces invariant or pairwise exchange compatible subspaces.
The results analogous to Eq.~\eqref{eqn:Toffdiag} for unitary and antiunitary antisymmetries are respectively:
\begin{subequations}
\begin{eqnarray}
W_{ij} H_j &=& - H_i W_{ij},
\label{eqn:Soffdiag} \\
W_{ij} H_j^{*} &=& - H_i W_{ij}.
\label{eqn:Poffdiag}
\end{eqnarray}
\end{subequations}

\subsection{Onsite symmetries of $\bk$-dependent Hamiltonians}

\co{Generalization to k-space of onsite unitary symmeties.}
The above methods extend to the onsite symmetries of $\bk$-space Hamiltonians of arbitrary dimensions.
An onsite unitary symmetry acts locally in $\bk$-space and is independent of $\bk$.
Given a family of Hamiltonians $H(\bk, \alpha)$, we treat linearly independent functions of $\bk$ as additional free parameters and apply the methods of section~\ref{sec:find_U}.

\co{Onsite antiunitaries.}
We now turn to time reversal symmetry, which requires special treatment because it transforms $\bk$ to $-\bk$.
Because $H(-\bk)$ is a reparametrization of the same Hamiltonian family, it is reduced if $H(\bk)$ is reduced.
The generalization of \eqref{eqn:Toffdiag} to the $\bk$-dependent case is
\begin{equation}
W_{ij} H_j^*(-\bk) = H_i(\bk) W_{ij}.
\end{equation}
By the same argument as before, separating the Hamiltonian to irreducible blocks guarantees that the nonzero solutions are unique and unitary up to constant factors.
The analogous results are true for particle-hole and chiral symmetry.

\subsection{Point group symmetries}
\co{Statement of the problem, real space rotation part is assumed given.}
The point group of a crystal is always a subgroup of the finite point group of its Bravais lattice.
Therefore we search for point group symmetries by enumerating possible real space rotations $R_g$, and applying methods similar to the previous subsections to find whether it is a symmetry with appropriate $U_g$.

\co{State analogous result.}
Like discrete onsite symmetries, point group symmetries may be combined with onsite unitaries, forming continuous families.
This ambiguity is again removed by using the reduced Hamiltonian.
The analogous result to \eqref{eqn:Toffdiag} for point group symmetries is $\left(U_g\right)_{ij} = \id \otimes W_{ij}$ where the blocks $W_{ij}$ satisfy
\begin{equation}
W_{ij} H_{j}(\bk) = H_{i}(R_g \bk) W_{ij}.
\label{eqn:PG}
\end{equation}
Here $W$ only has one nonzero block per row and column, and nonzero blocks only between compatible subspaces.
If the order of the symmetry is greater than 2, permutations which are not symmetric are also possible.
Because both $H_j(\bk)$ and $H_i(R_g \bk)$ are reduced, the nonzero solution for $W_{ij}$ is unique and unitary up to normalization and a phase factor.
With the knowledge of the full point group, the arbitrary phase factors appearing in $W_{ij}$ may be fixed such that the $U_g$ form a (double)group representation of the point group.

\co{Antiunitary and antisymmetry cases.}
A similar argument applies to the case of antiunitary point group symmetries (magnetic group symmetries) and antisymmetries that involve spatial transformations.
The analogous equations for unitary antisymmetries antiunitary (anti)symmetries are respectively:
\begin{subequations}
\begin{eqnarray}
W_{ij} H_{j}(\bk) &=& - H_{i}(R_g \bk) W_{ij},\\
W_{ij} H^*_{j}(-\bk) &=& \pm H_{i}(R_g \bk) W_{ij}.
\end{eqnarray}
\end{subequations}

\subsection{Continuous rotations}

\co{Continuous rotations are simpler and require less information.}
To find continuous rotation symmetries of $\bk\cdot\bp$ Hamiltonians we utilize the symmetry-adapted basis of the onsite unitary symmetries again.
Unlike discrete symmetries, the unitary action of a continuous symmetry cannot mix different blocks, because it continuously deforms to the identity.
Therefore we treat reduced Hamiltonians $H_i$ separately.

\co{Formal equation to solve.}
In order to find a continuous symmetry generator $g'$ as defined in sec.~\ref{sec:sym_constraints}, we simultaneously solve
\begin{equation}
\label{eq:find_contr}
g'(H_j)(\bk) = i [H_j(\bk), L^{(j)}] + i \sum_{lp}\frac{\partial H_j}{\partial k_l} M_{lp} k_p = 0
\end{equation}
for every $j$, with constraints $L^{(j)} = (L^{(j)})^{\dag}$, $\Tr L^{(j)} = 0$ and $M = - M^T = M^{\dag}$.
We then expand $H_j(\bk)$ in a basis of monomials, and reduce \eqref{eq:find_contr} to a system of linear equations for the entries of $L^{(j)}$ and $M$.

\section{Applications}
\co{Implementation}
We implemented the symmetric Hamiltonian generator and symmetry finder algorithms of the previous sections in the Qsymm Python package\cite{supp_mat, supp_mat_2}.
We provide an interface to define symbolic expressions of symmetries and Hamiltonian families using Sympy\cite{Meurer2017} and Kwant\cite{Groth2014}.
Efficient solving of large systems of linear equations is achieved using ARPACK\cite{Lehoucq1997} (bundled for Python by Scipy\cite{Jones2001}).
We provide the the source code with instructive examples as a software repository\cite{supp_mat, supp_mat_2}.
The following examples illustrate how the algorithms were used to solve open research problems in condensed matter physics.
We also provide the Jupyter notebooks\cite{Kluyver2016} generating these results.

\subsection{Symmetries of Majorana wire}
\co{Refer to paper.}
An early version of our symmetry finding algorithm was used in Ref.~[\onlinecite{Nijholt2016}] to find the symmetries of a superconducting nanowire in an external magnetic field.
The analysis revealed unexpected symmetries of the model system in certain geometries that prevent band tilting and closing of the topological gap.
Here we revisit these results.

\co{introduce Majoranas in general.}
The system under consideration is an infinite nanowire along the $x$ axis with a semiconducting core and a superconducting shell covering some of the surface.
In the presence of a magnetic field along the wire and a normal electric field, the wire undergoes a topological phase transition.
This is marked by the gap closing and reopening, with Majorana zero modes appearing at each end of a finite wire segment \cite{Leijnse2012}.
A component of the external magnetic field normal to the wire axis breaks the symmetry of the band structure ($E(\bk)\neq E(-\bk)$), leading to tilting of the bands and closing of the superconducting gap at finite momentum.
\begin{figure}[!tbh]
\includegraphics[width=0.97\columnwidth]{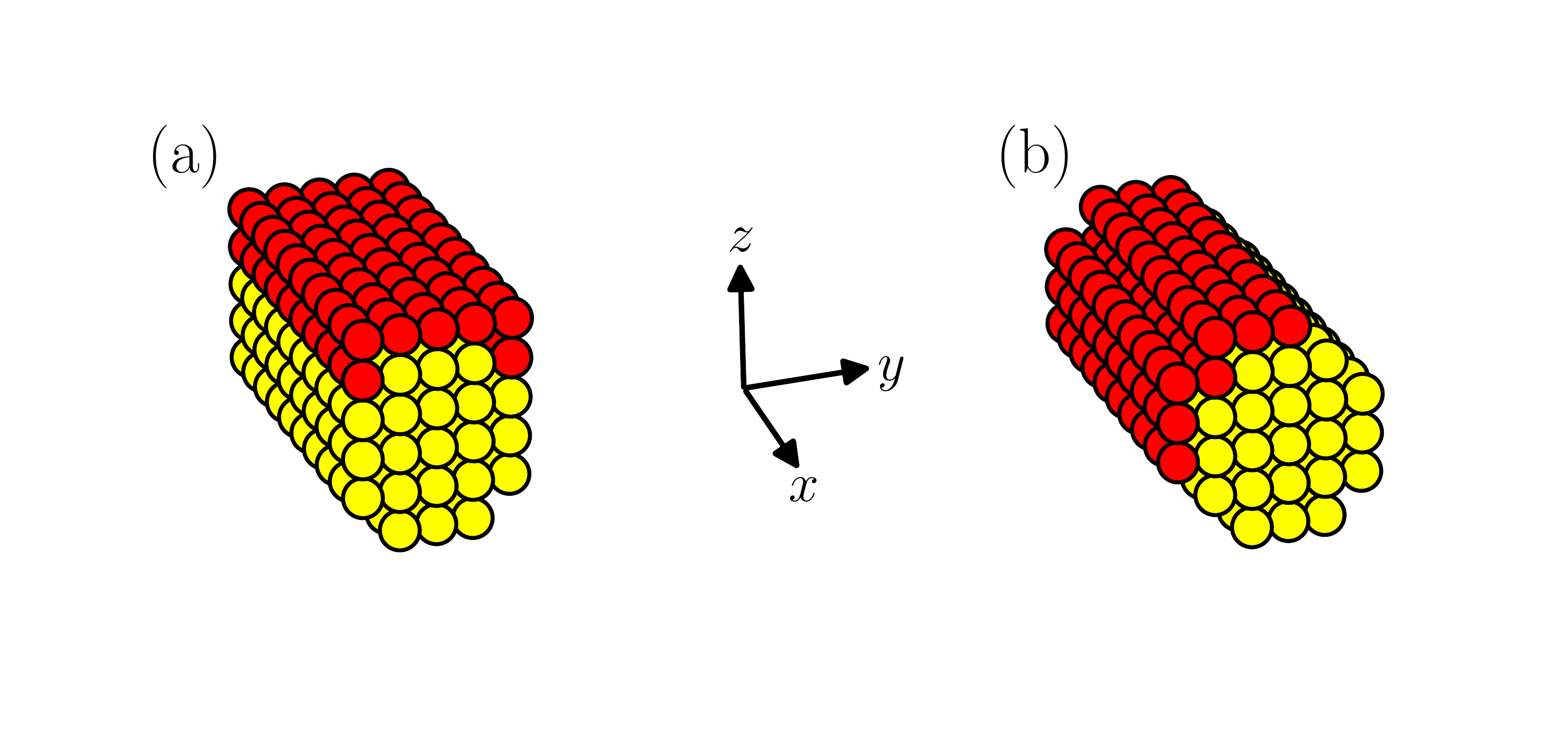}
\caption{Sketch of two possible geometries for the nanowire (yellow) with a superconducting shell (red). In (a), the geometry respects the mirror symmetry $M_y$, which gives rise to a chiral symmetry. In (b) however, the superconducting shell breaks the mirror symmetry, and hence the chiral symmetry is also absent.
Although the sketches show finite segments, the wire is translationally invariant along its axis $x$ in both cases.}
\label{fig:wires}
\end{figure}

\co{Discuss symmetries of the wire in the high symmetry case.}
Reference~[\onlinecite{Nijholt2016}] studied a tight-binding Hamiltonian of the wire and observed that depending on the geometry the band tilting may or may not occur.
Applying the symmetry finder algorithm to wires with small cross-sections (Fig.~\ref{fig:wires}), we identify the key difference in symmetry.
If the wire geometry has a mirror plane including its axis ($M_y$) as in Fig.~\ref{fig:wires}(a), with external fields $\mathbf{E} \parallel z$, and $\mathbf{B} \parallel x$, we find that the symmetry group consists of 8 elements.
The three generators of this group are particle-hole symmetry ($\cP$), a mirror plane perpendicular to the wire axis ($M_x$) and the combination of $M_y$ with time reversal $\cT$.
This last symmetry $M_y \cT$ includes both a spatial transformation and time reversal, and is easily overlooked.
This operator can be further combined with particle-hole to result in a chiral symmetry $\cC' = M_y \cT\cP$, as pointed out in the earlier work.
A nanowire enhanced by such an effective time reversal symmetry belongs to class BDI and supports multiple Majorana modes at its end\cite{Li2014}.

\co{Various symmetry breakings.}
Symmetries $M_x$ and $M_y \cT$ require $E(k_x) = E(-k_x)$ and prevent band tilting.
Adding nonzero $B_z$ to the magnetic field breaks $M_x$, but preserves $M_y \cT$, still forbidding band tilting.
Further reducing the symmetry by moving the position of the superconducting cover as in Fig.~\ref{fig:wires}(b), or by applying $B_y$ breaks all symmetries relating $k_x$ to $-k_x$ and enables band tilting.

\subsection{Kekule distortion in graphene}
\begin{figure}[t]
\begin{center}
\includegraphics[width=8.5cm]{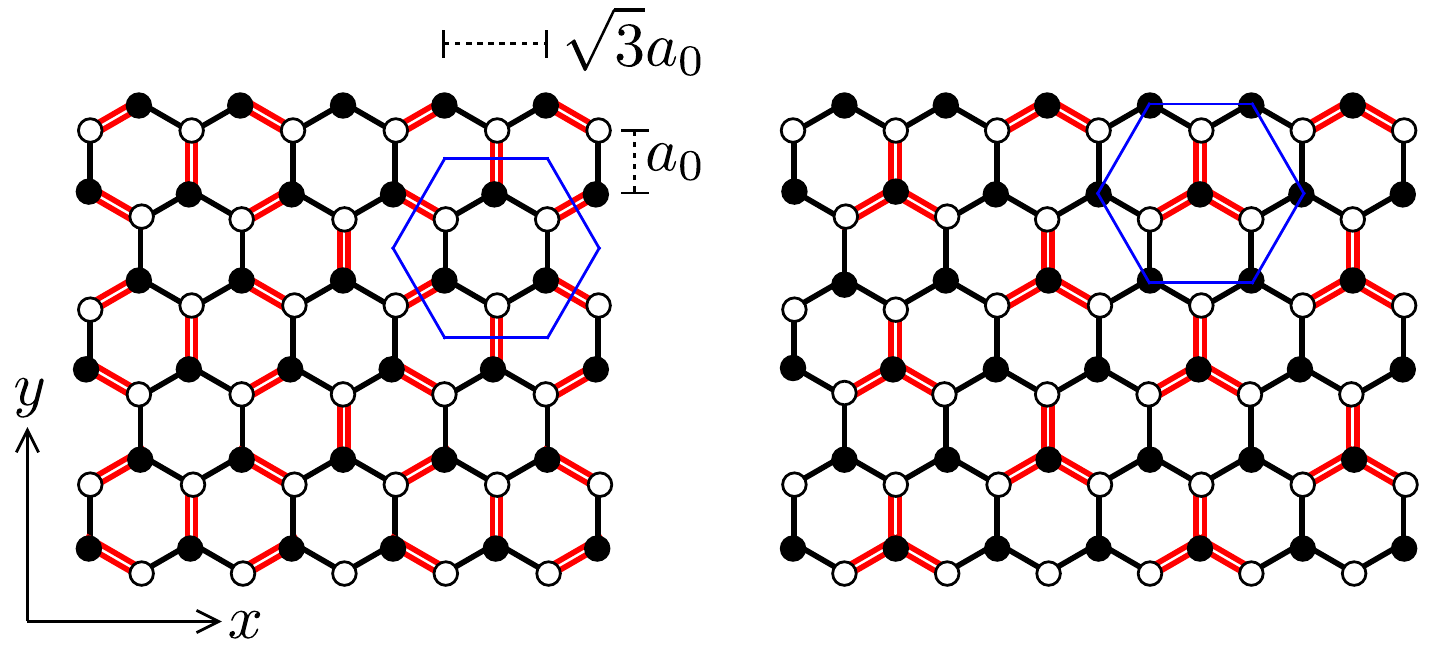}
\caption{Lattice structures of Kekule-O (left) and Kekule-Y patterns (right). Weak and strong bonds are marked with single and double lines. The high symmetry unit cell is marked in blue.}
\label{fig:Kekule}
\end{center}
\end{figure}

\co{Generalities about Kekule distortion.}
The Kekule distortion of graphene is a periodic pattern of weak and strong bonds tripling the size of the unit cell.
In the Kekule-O pattern, weak bonds around plaquettes resemble benzene rings, and in Kekule-Y, strong bonds form Y shapes around sites (Fig.~\ref{fig:Kekule}).
After folding back the Brillouin zone, the $K$ and $K'$ points are both mapped to the $\Gamma$ point.
A suitable mass term can now open a gap in the band structure.
A recent work\cite{Gamayun2018} reported that unlike the Kekule-O distortion\cite{Chamon2007}, the Kekule-Y distortion does not open a gap: instead it preserves a double Dirac cone at the Brillouin zone center.
Using our algorithms we identify the symmetries protecting this double Dirac cone.
\co{Symmetries of the continuum model of Kek-Y, sublattice symmetry is needed to protect double Dirac cone.}
First we find all the symmetries of the effective 4-band $\bk\cdot\bp$ model of Kekule-Y:
\begin{equation}
H_Y = v_1 (k_x \sigma_x + k_y \sigma_y) + v_2 (k_x \tau_x + k_y \tau_y),
\end{equation}
with $v_1, v_2$ band structure parameters.
We find that it is symmetric under the full hexagonal point group $D_6$ (in fact the linearized model has a continuous rotation symmetry), time reversal and sublattice symmetry, which results from the bipartite nature of the honeycomb lattice.
Next, we systematically generate all subgroups of this symmetry group and the corresponding symmetry-allowed 4-band $\bk\cdot\bp$ Hamiltonians.
We find that at least one antisymmetry is required to forbid a constant mass term that would open a gap at $\Gamma$.
A minimal subgroup protecting the double Dirac cone is generated by sublattice symmetry and three-fold rotations.
Removing sublattice symmetry, even while keeping the full $D_6$ point group, removes the protection of the double Dirac cone.
Sublattice symmetry is broken by adding second neighbour hopping, or a staggered onsite potential compatible with the lattice symmetries in the tight-binding model.

\co{Differences in the continuous rotation operators in the two cases.}
Symmetry finding shows that the effective model of Kekule-O,
\begin{equation}
H_O = v (k_x \sigma_x + k_y \sigma_y) + \Delta \sigma_z \tau_x
\end{equation}
with $v, \Delta \in \mathbb{R}$, has the same symmetry group structure.
However, the mass term $\sigma_z \tau_x$ is allowed even in the presence of the full symmetry group.
The key difference is the unitary action of rotations: the generator of continuous rotations is $\sigma_z + \tau_z$ in the Kekule-Y case, while it is $\sigma_z$ in the Kekule-O case.
Sublattice symmetry is $\cC = \sigma_z \tau_z$ in both cases.
Therefore no constant matrix can simultaneously anticommute with $\cC$ and commute with the rotation generator in Kekule-Y, while a mass term is allowed in Kekule-O.

\co{Cause of the difference from lattice model.}
This difference in the transformation properties stems from the different Wyckoff positions of the lattice sites.
In Kekule-Y the three-fold rotation centers lie on lattice sites, while in Kekule-O the three-fold rotation centers lie at centers of hexagonal plaquettes.
Using the tight-binding Hamiltonian generator, we confirm that the representation of 3-fold rotations in the low energy subspace at the center of the Brillouin zone is different for the Kekule-O and Kekule-Y systems.

\subsection{$\bk\cdot\bp$ model of distorted SnTe}
\co{Refer to Alex's paper on Weyl points and nodal lines in SnTe}
The cubic rocksalt material SnTe is the first example of topological crystalline insulators\cite{Hsieh2012}.
Recently, using our method, Ref.~\onlinecite{Lau2018} proposed that structural distortions can give rise to Weyl and nodal-line semimetal phases in the same material.
Here we review these results.

\co{Symmetries of the undistorted model}
The band gap of the cubic phase is smallest at the $L$ point in the face-centered cubic Brillouin zone.
We construct an effective $\bk\cdot\bp$ model expanded up to second order in $\bk$ around $L$.
The model has two orbital degrees of freedom, spanned by $p$ orbitals on Sn and Te sites.
The initial symmetry group of the $L$ point is $D_{3d}$ which is generated by inversion $I$, a three-fold rotation $C_3$ about the $\Gamma L$ axis, and a reflection $M_x$ about the mirror plane containing $\Gamma$ and two $L$ points.
Furthermore, the model should be invariant under time reversal $\cT$.
The corresponding representations of the symmetry operators, listing the (anti)unitary action first and the $\bk$-space action second, are as follows,
\begin{subequations}
\begin{align}
M_x &= -i s_x, & k_x &\to -k_x,\\
C_3 &= e^{i \frac{\phi}{2} s_z}, & \bk &\to
\begin{bmatrix}
\cos\phi & -\sin\phi & 0 \\
\sin\phi & \cos\phi & 0 \\
0 & 0 & 1
\end{bmatrix} \bk, \\
I &= \sigma_z, & \bk &\to -\bk, \\
\cT &= i s_y \cK, & \bk &\to -\bk,
\end{align}
\end{subequations}
where $\phi = \frac{2\pi}{3}$, and Pauli matrices $\sigma_i$ and $s_i$ act on orbital and spin degrees of freedom respectively. $\bk$ is the momentum vector measured from an $L$ point in a coordinate system where the $z$ axis is aligned with $\Gamma L$ (\emph{e.g.} [111]) and the $x$ axis is normal to a mirror plane (\emph{e.g.} [1$\overline{1}$0]).
The $\sigma_z$ in the unitary action of inversion is a result of considering an $L$ point, because the inversion center is one of the sites in the unit cell of the rocksalt structure, the other site is translated by a lattice vector under inversion and acquires a phase factor at nonzero momentum.

\co{Results of Hamiltonian generation with various symmetries}
Applying the $\bk\cdot\bp$ Hamiltonian generator algorithm we find 8 symmetry-allowed terms.
Ignoring the 3 terms that are proportional to the identity and do not influence band topology, we obtain the following Hamiltonian family:
\begin{align}
H_0(\bk) = & m\sigma_z + \nu (k_x s_y - k_y s_x) \sigma_x + \nu_3 k_z \sigma_y \nonumber\\
&+ c k^2_z \sigma_z + f(k^2_x + k^2_y)\sigma_z.
\end{align}
Breaking the three-fold rotational symmetry results in 8 new terms, 6 of which are not proportional to the identity:
\begin{align}
H_1(\bk) = \delta\nu (k_x s_y + k_y s_x) \sigma_x + \lambda_1 k_x s_z \sigma_x + \lambda_2 k_y \sigma_y \nonumber\\
 + \lambda_3 k_z s_x \sigma_x + \delta f(k_x^2 - k_y^2) \sigma_z + g k_y k_z \sigma_z.
\end{align}
Further breaking inversion symmetry produces 22 additional terms, none of which is proportional to the identity.

\subsection{Three-orbital tight-binding model for monolayer transition metal dichalcogenides}
\co{Why monolayer transition metal dichalcogenides are interesting.}
Monolayers of transition metal dichalcogenides MX$_2$ (M = Mo, W; X = S, Se, Te) are promising materials for use in electronics and optoelectronics \cite{Wang2012}.
When doped, the MX$_x$ monolayers also become superconducting \cite{Ye2012}.
In the $1H$ stacking, a monolayer consists of a layer of transition metal atoms M sandwiched between two layers of chalcogen atoms X.
Each layer separately is a triangular Bravais lattice, with the X atoms in the top and bottom layers projecting onto the same position in the plane of M atoms, forming an overall honeycomb lattice.
In the normal state, the monolayer is a semiconductor, with conduction and valence band edges at the corners of the hexagonal Brillouin zone $\pm K$.
Using that the wave functions at the band edges are predominantly composed of $d$-orbitals on the M atoms, Liu \emph{et al.}~[\onlinecite{Liu2013}] developed a three-orbital tight-binding model with nearest neighbor hopping.
This model satisfies the symmetry group of the monolayer, and has band edges near $\pm K$.
Here, we reproduce their spinless tight-binding model using our algorithm for symmetric Hamiltonian generation.

\co{Generate 3-band tight-binding model.}
The tight-binding basis consists of three $d$-orbitals on the M atom, namely
\begin{equation}
\psi = [|d_{z^2} \rangle, |d_{xy} \rangle, |d_{x^2-y^2} \rangle ]^T.
\end{equation}
Because the model does not include any orbitals on the X atoms, it has a triangular lattice, with lattice vectors ${\bf a}_1 = \hat{x}$ and ${\bf a}_2 = (\hat{x} + \sqrt{3} \hat{y})/2$.
The symmetry generators are time reversal symmetry $\mathcal{T}$, mirror symmetry $M_x$, and three-fold rotation in the monolayer plane $C_3$ which are represented in the tight-binding basis as
\begin{subequations}
\begin{align}
M_x &= \mathrm{diag}(1, -1, 1), & k_x &\to -k_x,\\
C_3 &= \begin{bmatrix}
1 & 0 & 0 \\
0 & \cos\phi & -\sin\phi \\
0 & \sin\phi & \cos\phi
\end{bmatrix},
& \bk &\to
\begin{bmatrix}
\cos\phi & -\sin\phi \\
\sin\phi & \cos\phi
\end{bmatrix} \bk, \\
\cT &= \cK, & \bk &\to -\bk,
\end{align}
\end{subequations}
with $\phi = \frac{2\pi}{3}$.
Employing the symmetrization strategy for lattice Hamiltonians described in Section \ref{section:gen_lattice}, we reproduce the tight-binding model of Ref.~\onlinecite{Liu2013}, given by
\begin{equation}
H(\bk) =
\begin{bmatrix}
h_{00}(\bk) & h_{01}(\bk) & h_{02}(\bk)\\
h_{01}^*(\bk) & h_{11}(\bk) & h_{12}(\bk) \\
h_{02}^*(\bk) & h_{12}^*(\bk) & h_{22}(\bk)
\end{bmatrix},
\end{equation}
with the matrix elements
\begin{eqnarray}
h_{00} & = & 2t_0 (2\cos \xi\cos\gamma+\cos2 \xi) + \epsilon_1  \notag \\
h_{01} & = & 2 i t_1 (\sin 2\xi + \sin\xi\cos\gamma) - 2\sqrt{3}t_2\sin\xi\sin\gamma, \notag \\
h_{02} & = & 2 i \sqrt{3} t_1 \cos\xi\sin\gamma + 2t_2(\cos 2\xi - \cos\xi\cos\gamma), \\
h_{11} & = & t_{3} (\cos\xi\cos\gamma + 2 \cos2 \xi) + 3 t_{4} \cos\xi \cos\gamma + \epsilon_2, \notag \\
h_{12} & = & \sqrt{3} (t_{4}-t_{3})\sin\xi\sin\gamma + 4 i t_{5} \sin\xi(\cos\xi - \cos\gamma), \notag \\
h_{22} & = & 3t_{3} \cos\xi\cos\gamma + t_{4}(\cos\xi\cos\gamma +2\cos2\xi )+\epsilon_2, \notag
\end{eqnarray}
where $\xi=k_x/2$ and $\gamma=\sqrt 3 k_y/2$ and the lattice constant is set to one.

\subsection{Lattice Hamiltonian of monolayer WTe$_2$}
\label{sec:WTe2}
\co{Why WTe2 is interesting.}
Monolayer WT2$_2$ was recently discovered to be a two-dimensional quantum spin Hall insulator\cite{Fei2017, Tang2017, Wu2018, Jia2017} in accordance with previous numerical prediction\cite{Qian2014, Zheng2016}.
Transport experiments found quantized edge conductivity persisting up to 100K\cite{Wu2018}.
This suggests a much larger band gap compared to devices based on two-dimensional quantum wells\cite{Konig2007}.
It remains an open question whether a simple non-interacting lattice Hamiltonian can reproduce these findings.

\co{How we generate TB Hamiltonian.}
We use the restricted set of orbitals in Ref.~\onlinecite{Muechler2016} to construct the spinless tight-binding Hamiltonian.
The unit cell contains four sites (labeled $A_d$, $A_p$, $B_d$, $B_p$) with one orbital on each, and has a symmetry group generated by time reversal, inversion and glide reflection.
We use the permutation of the sites under the symmetries and the onsite unitary action (in this case $\pm 1$ factors) as input.
The model includes hoppings of type $A_p$--$A_p$, $B_d$--$B_d$ in neighboring unit cells in the $x$ direction, and $B_d$--$A_p$, $A_p$--$B_p$ and $A_d$--$B_d$ within the unit cell.
We reproduce the Hamiltonian family with 7 free parameters also found in the reference:
\begin{widetext}
\begin{equation}
H(\bk) = \begin{bmatrix} \mu_{d} +  2 t_{d} \cos k_{x} & 0 &  2 t_{d}^{AB}  f_d (\bk) & 2 i t_{0}^{AB} g(\bk)\\
0 &  \mu_{p} + 2 t_{p} \cos k_x &  -2 i t_{0}^{AB} g(\bk) &  2 t_{p}^{AB} f_p (\bk)\\
\textnormal{h.c.} & \textnormal{h.c.} &  \mu_{d} +  2 t_{d}  \cos k_{x} & 0\\
\textnormal{h.c.} &  \textnormal{h.c.} & 0 & \mu_{p} +  2 t_{p} \cos k_{x}
\end{bmatrix},
\end{equation}
\end{widetext}
where
\begin{eqnarray}
f_l (\bk) &=& \cos( k_{x} x_{Al} - k_{x} x_{Bl}) e^{i k_{y} y_{Al} + i k_{y} y_{Bl}},\\
g(\bk) &=& \sin (k_{x} x_{Ap} - k_{x} x_{Bd}) e^{- i k_{y} y_{Ap} + i k_{y} y_{Bd}},
\end{eqnarray}
for $l\in [p, d]$ and the lattice vectors are $[1, 0]$ and $[0, 1]$.
This Hamiltonian is identical to the one found previously up to transformations of the Bloch basis.

\co{Advertise result with SoC.}
Extending this analysis to include spin and possible spin-orbit coupling terms, we find that there are 7 additional terms allowed by symmetry in a tight-binding model with the same bonds.
The detailed results will be published elsewhere\cite{Lau2018b}.

\section{Summary}
\co{Summarize results.}
Analysis of condensed matter systems is commonly based on single-particle Hamiltonians, the symmetry properties and classification of which are crucial to understanding the physical properties.
We discussed the general symmetry structure of single-particle Hamiltonian families, and presented methods to find the full symmetry group of a Hamiltonian, and to generate all Hamiltonians compatible with a given symmetry group.
Our methods extend to all continuous and discrete symmetries of single-particle continuum or lattice Hamiltonians.

Although we focused on fermionic systems, the framework we presented is generally applicable whenever the form of the symmetry action and the Hamiltonians is the same, e.g.~in the analysis of unconventional superconducting pairing, or even Josephson junction arrays.
Our algorithms provide a powerful tool in the ongoing classification of symmetry protected topological phases in a wide variety of physical settings ranging from classical mechanics to circuit QED.
The Hamiltonian generator can be extended to search for nonlinear effective field theories and interacting lattice models respecting given symmetries.
The symmetry finder may also be further generalized to facilitate more involved symmetry analysis by decomposing group representations.
We leave these open questions to future work.

\co{Algorithms and implementation.}
We implemented the algorithms in the Qsymm Python package, making them easily accessible.
We demonstrated the usefulness of our approach by applying it to a number of relevant modern research topics including graphene, transition metal dichalcogenides and topological semimetals, resulting in several new insights.

\acknowledgments
We thank N. V. Gnezdilov, M. Hastings, A. Lau, B. Nijholt, V. P. Ostroukh, R. Skolasinski and J. B. Weston for fruitful discussions.
This work was supported by ERC Starting Grant 638760, the Netherlands Organisation for Scientific Research (NWO/OCW), and the US Office of Naval Research.

\emph{Author contributions:} A. Akhmerov provided the initial idea and oversaw the project.
T. O. Rosdahl developed the algorithm to constrain Hamiltonians and the early version of the symmetry finding algorithm.
D. Varjas developed the tight-binding Hamiltonian generator and general symmetry finding algorithms.
T. O. Rosdahl and D. Varjas wrote the source code of the Qsymm package with the supervision of A. Akhmerov.
All authors contributed to writing the manuscript.

\bibliographystyle{apsrev4-1}
\bibliography{bibl}

\appendix

\section{Simultaneous diagonalization}
\label{sec:simult_diag}
We present an algorithm to simultaneously diagonalize a set of mutually commuting normal matrices.
The key property that follows from the commutation is that the matrices share eigensubspaces.
By transforming to the diagonalizing basis of one of the matrices, the rest of the matrices are guaranteed to be block diagonal with blocks corresponding to the degenerate eigensubspaces of the first.
Considering this, we apply the following recursive algorithm to find the simultaneous eigenvectors spanning the simultaneous eigensubspaces of commuting matrices $H_i$:
\begin{itemize}
\item If the matrices are $1\times 1$, return the $1\times 1$ identity matrix.
\item Diagonalize the first matrix $H_0$, find the orthonormal sets of eigenvectors spanning each (approximately) degenerate eigensubspace.
This results in a set of projectors $P_j$ onto the eigensubspaces, each consisting of a set of orthonormal column vectors, the number of columns equal to the degeneracy of the $j$'th eigensubspace.
\item If there are no more matrices, return this basis.
\item Project the rest of the matrices into each degenerate eigensubspace $j$: $\tilde{H}_{ij} = P_j^{\dag} H_i P_j$ for $i > 0$.
\item Perform this algorithm on the projected matrices $\tilde{H}_{ij}$ ($i>0$) in each eigensubspace $j$, this returns a set of projectors $\tilde{P}_{jk}$.
\item Return the set of projectors $P_{jk} = P_j \tilde{P}_{jk}$ for every $j$ and $k$.
\end{itemize}
The output is a set of projectors $P_i$, each consisting of a set of orthonormal column vectors spanning a simultaneous eigensubspace of the $H_i$'s.
Horizontally stacking the $P_i$'s gives a unitary matrix $U$ such that $U^{\dag} H_i U$ is diagonal for all $i$.
The algorithm is guaranteed to finish, as at each recursion level both the number and the size of the matrices is decreased.
The main source of numerical instability is the decision whether to treat two numerically close eigenvalues as degenerate or not.
The algorithm is most stable if matrices that have eigenvalues which are either well separated or degenerate to machine precision are first and those which may have accidental near-degeneracies are last.
In the physical problems we consider symmetry operators and projectors are of the first kind, while Hamiltonians are of the second.

\section{Finding the symmetry-adapted basis}
\label{sec:sym_bas}
Our goal in this section is to find the symmetry-adapted basis on a Hilbert space of dimension $n d$.
We have already ensured that the algebra of conserved quantities forms a representation of $\mathfrak{su}(d)$, such that in the proper basis the generators have the tensor product structure of $L \otimes \id_{n\times n}$ where the matrices $L$ span the space of all traceless $d\times d$ Hermitian matrices.

We pick a generator $L\in \mathfrak{su}(d)$.
Given the tensor product structure, every eigenvalue of the generators has degenaracy which is a multiple of $n$.
In the case when some eigenvalues of $L$ have degeneracy higher than $n$, we restrict the other generators to the $fn$-dimensional (with $f \leq d$ integer) eigensubspace, where they span $\mathfrak{su}(f)$.
By iterating over the other generators in this restricted space it is always possible to find one that has eigenvalue degeneracy lower than $fn$, until all degeneracies larger than $n$ are split.
This procedure results in a basis with $n$-dimensional subspaces grouped together, but their bases not aligned with each other.

In this basis $L$ has the diagonal form
\begin{equation}
L = \begin{bmatrix}
 L_{11} \id_{n\times n}      & 0  &  \cdots \\
 0  &  L_{22} \id_{n\times n}  & \cdots \\
\vdots      &  \vdots     & \ddots
\end{bmatrix}.
\label{eqn:blockL2}
\end{equation}
We wrote the matrix in a different block form compared to (\ref{eqn:blockL}), in this notation the symmetry adopted basis is characterized by every block being proportional to $\id_{n\times n}$ for every element $M$ of $\mathfrak{g}$:
\begin{equation}
\tilde{M} = \begin{bmatrix}
 M_{11} \id_{n_i\times n_i}      & M_{12} \id_{n_i\times n_i}   &  \cdots \\
 M_{21} \id_{n_i\times n_i}   &  M_{22} \id_{n_i\times n_i}  & \cdots \\
\vdots      &  \vdots     & \ddots
\end{bmatrix}.
\end{equation}
We know that such basis exists with a selected generator $L$ having the diagonal form (\ref{eqn:blockL2}).
Every unitary basis transformation preserving (\ref{eqn:blockL2}) has the block diagonal form
\begin{equation}
U = \begin{bmatrix}
 U_1      & 0  &  \cdots \\
 0  &  U_2  & \cdots \\
\vdots      &  \vdots     & \ddots
\end{bmatrix}
\end{equation}
with $n\times n$ unitaries $U_k$. In this transformed basis $M$ reads
\begin{equation}
M = U \tilde{M} U^{\dag} = \begin{bmatrix}
 M_{11} \id_{n\times n}      & M_{12} U_1 U_2^{\dag}  &  \cdots \\
 M_{21} U_2 U_1^{\dag}   &  M_{22} \id_{n\times n}  & \cdots \\
\vdots      &  \vdots     & \ddots
\end{bmatrix}.
\end{equation}
By fixing $U_1 = \id_{n\times n}$ we can iterate over the nonzero off-diagonal blocks of $M$ and successively fix the basis for each block such that $U_i U_j^{\dag} = \id$.
It is always possible to find a generator in $\mathfrak{su}(d)$ that does not have a zero block in a given position in the diagonalizing basis of $L$.

By this procedure we find a symmetry-adapted basis where every generator has the tensor product structure $L \otimes \id_{n\times n}$ and the Hamiltonian commuting with these generators the structure $\id_{d\times d} \otimes H_r$ with $H_r$ the reduced Hamiltonian.
This structure is invariant under any unitary basis transformation $U \otimes V$ with $U \in U(d)$ and $V \in U(n)$, this is the ambiguity in the symmetry-adapted basis.

\section{Lemma on tensor product operators}
\label{app:tensor}

\begin{lemma}
Let $V_1\otimes V_2$ and $W_1 \otimes W_2$ be complex finite-dimensional tensor product Hilbert spaces. Define two families of operators that span the space of all linear operators on the first component and leave the second component of the tensor product invariant
\begin{eqnarray}
\mathfrak{g} &=& \brac{L_1\otimes \id_{V_2} | L_1 \in \mathcal{L}(V_1)} \\
\mathfrak{h} &=& \brac{M_1\otimes \id_{W_2} | M_1 \in \mathcal{L}(W_1)}.
\end{eqnarray}
Let $U: V_1\otimes V_2 \to W_1 \otimes W_2$ be a linear map that maps the two operator spaces into each other
\begin{eqnarray}
U \mathfrak{g} U^{\dag} &\subseteq & \mathfrak{h}\nonumber\\
U^{\dag} \mathfrak{h} U &\subseteq & \mathfrak{g}.
\label{eqn:Ugh}
\end{eqnarray}
Then $U$ has a tensor product form $U = U_1 \otimes U_2$ with $U_1: V_1 \to W_1$ and $U_2: V_2 \to W_2$. $U$ is nonzero only if $\dim V_2 = \dim W_2$, in this case $U_2$ can always be chosen unitary.
\end{lemma}

If $\mathfrak{g=h}$ and $U$ is unitary, the conditions (\ref{eqn:Ugh}) are equivalent to $U \mathfrak{g} U^{\dag} = \mathfrak{g}$.

To prove the lemma, we utilize the singular value decomposition (SVD) of $U$, treating it as a linear operator $V_1 \otimes W_1 \to V_2 \otimes W_2$:
\begin{equation}
U = \sum_{\alpha} s_{\alpha} U^{\alpha}_1 \otimes U^{\alpha}_2
\end{equation}
where $s_{\alpha}\in\mathbb{R}^+$ are the singular values, $U^{\alpha}_1$ and $U^{\alpha}_2$ nonzero vectors in $V_1 \otimes W_1$ and $V_2 \otimes W_2$ respectively.
Alternatively we can view them as linear maps $U^{\alpha}_1: V_1 \to W_1$, $U^{\alpha}_2: V_2 \to W_2$.
They satisfy the orthogonality condition $\Tr\pars{U^{\alpha}_1 (U^{\beta}_1)^{\dag}} \propto \delta^{\alpha\beta}$, $\Tr\pars{U^{\alpha}_2 (U^{\beta}_2)^{\dag}} \propto \delta^{\alpha\beta}$.

\begin{widetext}
We rewrite (\ref{eqn:Ugh}) as
\begin{eqnarray}
\forall L_1 \in GL(V_1) \:\exists M_1 \in GL(W_1) :\\ \nonumber
 U \pars{L_1\otimes \id_{V_2}} U^{\dag} =& \sum_{\alpha\beta} s_{\alpha} s_{\beta} \pars{U^{\alpha}_1 L_1 (U^{\beta}_1)^{\dag}}\otimes \pars{U^{\alpha}_2 (U^{\beta}_2)^{\dag}} &= M_1 \otimes \id_{W_2} \\
\forall M_1 \in GL(W_1) \:\exists L_1 \in GL(V_1) :\\ \nonumber
 U^{\dag} \pars{M_1\otimes \id_{W_2}} U =& \sum_{\alpha\beta} s_{\alpha} s_{\beta} \pars{(U^{\alpha}_1)^{\dag} M_1 U^{\beta}_1}\otimes \pars{(U^{\alpha}_2)^{\dag} U^{\beta}_2}  &= L_1 \otimes \id_{V_2}
\end{eqnarray}
\end{widetext}
The term $U^{\alpha}_1 L_1 (U^{\beta}_1)^{\dag}$ cannot be zero for every $L_1$ as it would imply that either $U^{\alpha}_1$ or $U^{\beta}_1$ vanishes.
$U^{\alpha}_1 L_1 (U^{\beta}_1)^{\dag} \propto U^{\alpha'}_1 L_1 (U^{\beta'}_1)^{\dag} \;\forall L_1$ iff $\alpha = \alpha'$ and $\beta = \beta'$, this is seen by treating  $U^{\alpha}_1 (.) (U^{\beta}_1)^{\dag}$ as linear operators $\mathcal{L}(V_1)\to \mathcal{L}(W_1)$ and using the orthogonality condition.
Analogous statements are true for $(U^{\alpha}_1)^{\dag} M_1 U^{\beta}_1$.
This restricts $U^{\alpha}_2 (U^{\beta}_2)^{\dag} \propto \id_{W_2}$ and $(U^{\alpha}_2)^{\dag} U^{\beta}_2 \propto \id_{V_2}$.
For $\alpha=\beta$ this means that $\operatorname{Ker} (U^{\alpha}_2)^{\dag} = 0$ and $\operatorname{Ker} U^{\alpha}_2 = 0$ which is only possible if $\dim V_2 = \dim W_2$ and $U^{\alpha}_2$ is invertible.
As $(U^{\alpha}_2)^{\dag} \propto (U^{\alpha}_2)^{-1}$, by moving a constant factor from $U^{\alpha}_2$ to $U^{\alpha}_1$ it is always possible to make $U^{\alpha}_2$ unitary. For $\alpha\neq\beta$, using the orthogonality condition we find $(U^{\alpha}_2)^{\dag} U^{\beta}_2 = 0$ which is impossible for invertible operators.
This shows that the SVD consists of a single term and concludes the proof.

\section{Proof of block structure of  symmetry operators}
\label{sec:disc_block}

We consider the general case of multiple irreps and show that an antiunitary (anti)symmetry takes a simple block structure in the symmetry-adapted basis. Explicitly writing the action $U L^* U^{\dag}$ of the unitary part of $\cT= U \cK$,
\begin{equation}
U = \begin{bmatrix}
 U_{11}      & U_{12}     &  \cdots \\
 U_{21}      &  U_{22}     & \cdots \\
\vdots      &  \vdots     & \ddots
\end{bmatrix}
\end{equation}
on the generic symmetry generator $L \in \mathfrak{g}$
\begin{equation}
L  = \begin{bmatrix}
L^{(1)} \otimes \id_{n_1\times n_1}     & 0     &  \cdots \\
 0     & L^{(2)} \otimes  \id_{n_2\times n_2}    & \cdots \\
\vdots      &  \vdots     & \ddots
\end{bmatrix}
\end{equation}
and demanding $U \mathfrak{g} U^{\dag} \subseteq	\mathfrak{g}$ and $U^{\dag} \mathfrak{g} U \subseteq	\mathfrak{g}$.
We find that
\begin{equation}
U_{ij} \mathfrak{g}_j U_{ij}^{\dag} \subseteq \mathfrak{g}_i \;\textnormal{and}\; U_{ij}^{\dag} \mathfrak{g}_i U_{ij} \subseteq	\mathfrak{g}_j
\end{equation}
where $\mathfrak{g}_i$ is the space of symmetry generators in block $i$, $\mathfrak{g_i} = \brac{L^{(i)}\otimes  \id_{n_i\times n_i}  }$.

By the lemma in Appendix~\ref{app:tensor}, $U_{ij} \neq 0$ only if $n_i = n_j$ and factorizes as $U_{ij} = V_{ij}\otimes W_{ij}$ with unitary $W_{ij}$.
Using this we also find that $U_{ki} \mathfrak{g}_i U_{ji}^{\dag} = 0$ ($k\neq j$), which means that either $U_{ki}$ or $U_{ji}$ vanishes, so there can be only one nonzero block in every row or column.
As $U U^{\dag} = \id$, each block needs to be unitary and the block structure of $U$ is restricted to that of a permutation matrix.
The determinant of such a matrix is only nonzero if the nonzero off-diagonal blocks are square: $U_{ij} \neq 0$ implies $n_i = n_j$ and $d_i = d_j$.
This allows $V_{ij}=\id$ to be chosen for all the nonzero blocks.

In the case of antiunitary (anti)symmetry $W W^*\propto \id$, if $W_{ij}$ is nonzero, $W_{ji}$ is also nonzero with $W_{ij} W_{ji}^* \propto \id$, the block structure of $W$ is restricted to that of a symmetric permutation matrix.
The analogous argument can be made in the case of unitary antisymmetries by dropping the complex conjugations.

\section{Beautification of Hamiltonian families and conserved quantities}
A Hamiltonian family \eqref{eq:family} is a linear space of Hamiltonians, and applying symmetry constraints to a family involves mapping the constraints to a generally rectangular matrix, such that the symmetry constrained subfamily of Hamiltonians lives in its null space [see \eqref{eq:linear_constraints}].
Numerically obtaining a basis for the null space, namely the symmetric subfamily, is straightforward using standard linear algebra methods.
However, numerical routines generally return basis vectors that are oriented along arbitrary directions in the subspace, and the resulting subfamily thus not necessarily as easily human readable as possible, containing many nonzero elements that are redundant.
To give a simple example, numerically computing a basis for a two-dimensional Euclidean plane might yield the vectors $[1/\sqrt{2}, \pm 1/\sqrt{2}]^T$, while the standard basis $\{[1, 0]^T, [0, 1]^T\}$ is more intuitive.

We take increased human readability of a Hamiltonian family to mean having a smaller number of nonzero elements in the matrix parts, and the span unchanged.
Since a family spans a linear space, we can express each family as a full rank matrix.
This is done by mapping each family member to a row vector by flattening and concatenating all the matrix coefficients, and vertically stacking these vectors.
Obtaining a human readable representation of the family then amounts to finding another matrix with the same row space but with as few nonzero entries as possible.
This problem is known in the literature as matrix sparsification \cite{Gottlieb2010}, and although widely studied, to our knowledge no general algorithms for matrix sparsification exist.

To solve this problem, we sparsify the matrix representation of a Hamiltonian family by bringing it to reduced row echelon form.
In reduced row echelon form, the first nonzero number from the left in a row is always equal to $1$, and is located to the right of the first nonzero entry in the row above.
Furthermore, the first nonzero entry per row is always the only nonzero entry in its column, and the number of nonzero entries thus minimal.
In addition, bringing a matrix to reduced row echelon form preserves its row span.
We obtain the reduced row echelon form by performing elementary row operations on the matrix representation of the family.
In floating point precision, this generally leads to numerical instability.
However, for the applications we consider, this is not a major obstacle, since the matrices we consider are typically small, and usually only contain nonzero elements that are of the order $1$, such that the distinction between zero and nonzero entries is unambiguous.

Conserved quantities, which also form a linear space spanned by a set of matrices, suffer from the same ambiguity.
We apply the same algorithm of bringing the matrix whose rows are the flattened generators of conserved quantities to reduced row echelon form in order to bring the generator set to a more human readable form.

\end{document}